\documentclass[journal,twocolumn]{IEEEtran}

\usepackage{amsmath,amsthm,amssymb,scrextend}
\usepackage{fancyhdr}
\usepackage{relsize}
\usepackage{epstopdf} 
\usepackage{floatrow}
\usepackage[font={small}]{caption}
\usepackage{multirow}
\usepackage{balance}
\usepackage[numbers,sort&compress,square]{natbib}
\usepackage{subfigure}
\usepackage{ifpdf}
\ifCLASSINFOpdf
   \usepackage[pdftex]{graphicx}
   \graphicspath{{./Figures/pdf/}}
   \DeclareGraphicsExtensions{.pdf}
\else
   \usepackage[dvips]{graphicx}
   \graphicspath{{./Figures/eps/}}
   \DeclareGraphicsExtensions{.eps}
\fi

\begin{document}

\pagenumbering{gobble} 

\title{Optimization of a Millimeter-Wave UAV-to-Ground Network in Urban Deployments}
\author{
\IEEEauthorblockN{Enass Hriba,\IEEEauthorrefmark{1}
Matthew C. Valenti,\IEEEauthorrefmark{1}
and Robert W. Heath, Jr.\IEEEauthorrefmark{2} } \\
\IEEEauthorrefmark{1}West Virginia University, Morgantown, WV, USA. \\
\IEEEauthorblockA{\IEEEauthorrefmark{2}North Carolina State University, Raleigh, NC, USA.}
\vspace{-0.6cm}
}

\maketitle
\thispagestyle{empty}


\begin{abstract}

An urban tactical wireless network is considered wherein the base stations are situated on unmanned aerial vehicles (UAVs) that provide connectivity to ground assets such as vehicles located on city streets.  The UAVs are assumed to be randomly deployed at a fixed height according to a two-dimensional point process.   Millimeter-wave (mmWave) frequencies are used to avail of large available bandwidths and spatial isolation due to beamforming.  In urban environments, mmWave signals are prone to blocking of the line-of-sight (LoS) by buildings.  While reflections are possible, the desire for consistent connectivity places a strong preference on the existence of an unblocked LoS path.   As such, the key performance metric considered in this paper is the connectivity probability, which is the probability of an unblocked LoS path to at least one UAV within some maximum transmission distance.   By leveraging tools from stochastic geometry, the connectivity probability is characterized as a function of the city type (e.g., urban, dense urban, suburban), density of UAVs (average number of UAVs per square km), and height of the UAVs.  The city streets are modeled as a Manhattan Poisson Line Process (MPLP) and the building heights are randomly distributed.   The analysis first finds the connectivity probability conditioned on a particular network realization (location of the UAVs) and then removes the conditioning to uncover the distribution of the connectivity; i.e., the fraction of network realizations that will fail to meet an outage threshold.  While related work has applied an MPLP to networks with a single UAV, the contributions of this paper are that it (1) considers networks of multiple UAVs, (2) characterizes the performance by a connectivity distribution, and (3) identifies the optimal altitude for the UAVs.

\end{abstract}

\section{Introduction}

Networks based on Unmanned Aerial Vehicles (UAVs) can provide instant infrastructure for services such as surveillance, broadband access, networked control, and automation \cite{Shang2019,Cicek2019,He20188,Kishk2020}.  For such networks, the use of millimeter-wave (mmWave) frequencies is an attractive option thanks to the high available bandwidth coupled with spatial isolation due to beamforming \cite{Rapp:2013,Liu2019}.  In urban areas, among the most challenging of environments, buildings pose a major source of blockage to the line-of-sight (LoS) propagation path \cite{Han2017,rappaport2014millimeter}.   While reflections are possible at mmWave, the desire for consistent and reliable connectivity places a strong preference on the existence of an unblocked LoS path. 

In this paper, we consider the performance and optimization of a network of randomly deployed UAVs in a typical urban environment.  The analysis leverages tools from stochastic geometry to model both the network realization (i.e., the UAV locations) and the city environment (i.e., the heights and locations of buildings).  The key performance metric developed here is the \emph{connectivity probability}, which characterizes the likelihood that a typical user of the network (assumed to be a vehicle located on a city street) is able to connect to a LoS UAV within some maximum communication range.

Stochastic geometry has previously been used to study the LoS probability in two-dimensional mmWave networks \cite{Venugopal2016,Andrews17,Hriba2017,Enass2018,Enass2019}, where the locations of blockages are drawn from an appropriate random process.  Those works are  not directed towards urban environments and do not account for the grid-like arrangements of city streets or the variable heights of buildings, not all of which will be tall enough to block the LoS.   The grid-like pattern of city streets is considered in \cite{BaccelliMPLP}, which uses a Manhattan Poisson Line Process (MPLP) to model patterns of perpendicular streets.   Building upon that work, \cite{Heath20211} also uses a MPLP to model the grid of streets, but it additionally accounts for random building heights.  The analysis of \cite{Heath20211} derives the LoS probability to a \emph{single} UAV as well as its spatial average (called the\emph{ area LoS probability}) when the UAV is uniformly distributed at fixed height within a circular cell.

Similar to \cite{BaccelliMPLP,Heath20211}, this paper uses a MPLP to account for the randomness of the city streets, and hence the random locations of buildings contained within blocks.   
Like \cite{Heath20211}, the random heights of buildings are considered, but unlike \cite{Heath20211}, this paper is not limited to a single UAV.  Hence, a main technical contribution of this paper is that it extends the results in \cite{Heath20211} to the case of a network of \emph{multiple} UAVs that are randomly scattered within the operational area.   
This extension allows for two key applications: (1) Quantification of the tradeoff between the UAV density (average number of UAVs per  $\mathsf{km^2}$) and the network performance, which identifies the minimum number of UAVs required to cover an area at a desired performance level, and (2) Identification of the optimal UAV height for a given deployment scenario (city type).


After introducing the system model in Section II, the paper provides an analysis in Section III.  The analysis begins by defining the connectivity probability, first considering the probability when it is conditioned on the placement of the UAVs, then removing the conditioning to reveal the \emph{distribution} of the connectivity; i.e., the probability that a network realization has a connectivity probability that is less than a threshold.   
By defining the $\emph{outage}$ probability to be the connectivity distribution evaluated at the minimum acceptable connectivity probability, we are able to characterize the tradeoff between the density of UAV  and outage probability. Finally, for a given UAV density and type of city (e.g., urban, dense urban, suburban), we are able to optimize the UAV height  by finding the value that minimizes the outage probability.  Numerical results are provided in Section IV, and the paper concludes with Section V.    



 \section{System Model}
\label{sec:Networktopology}
\begin{figure}[t]
		\centering
\includegraphics[width=.95\linewidth] {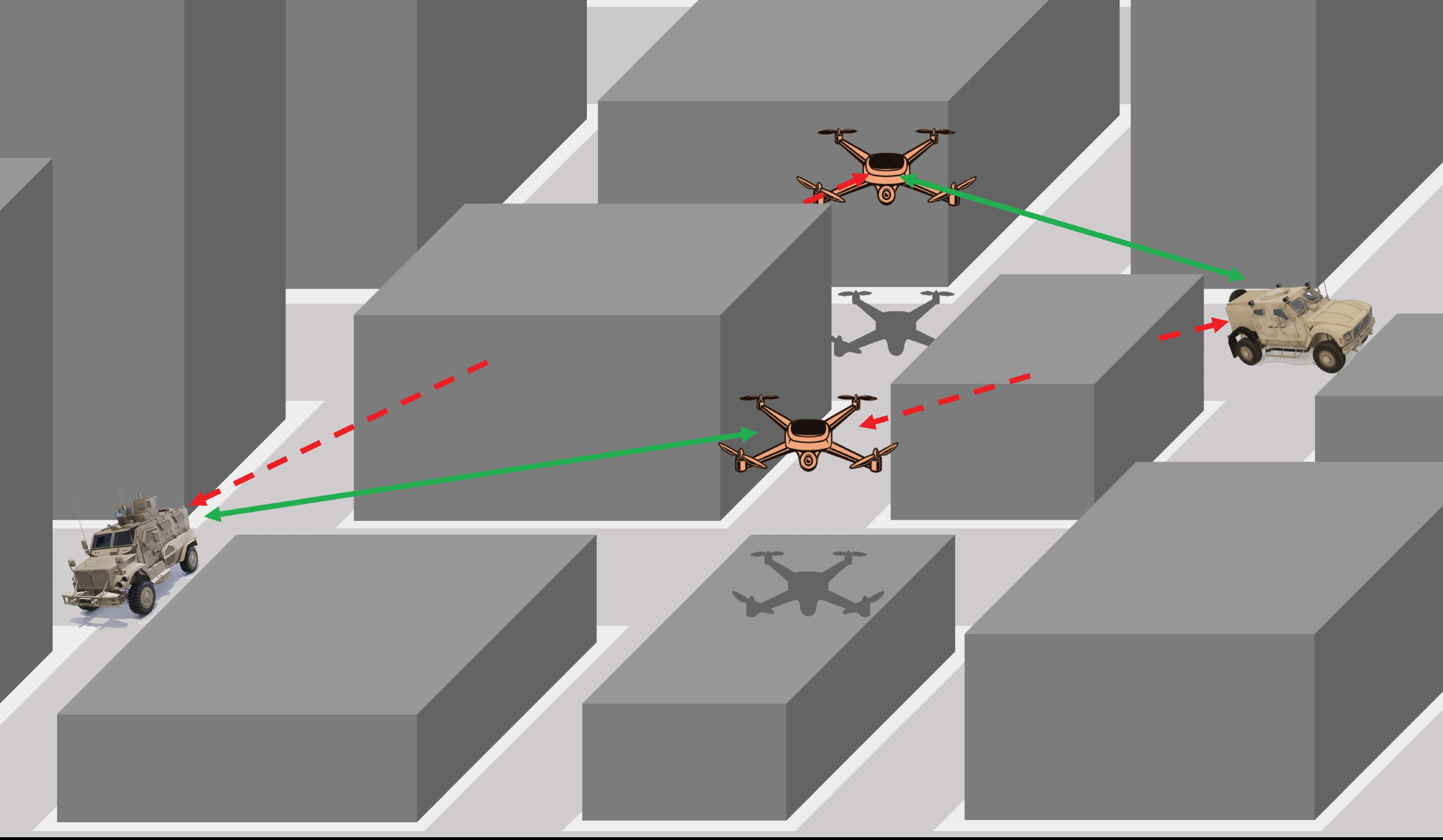}
	    \caption{Example of a 3D Urban aerial network consisting of UAVs, user vehicles, and buildings.  Green solid lines indicate LoS links while red dashed lines indicate non-LoS.  A shadow is shown directly below each UAV to identify its projection onto the streets or building tops. Street widths are exaggerated relative to building widths. \vspace{-0.5cm} }
		\label{fig:Model3D2}
	\end{figure}

As illustrated in Fig.~\ref{fig:Model3D2}, the network consists of UAVs in the air, users on the ground (typically located in vehicles), and blocks of buildings that serve as potential sources of blocking.   The figure shows two UAVs and two user vehicles. A shadow is shown directly below each UAV to identify its projection onto the building or street below it.  Each UAV is LOS to one of the UAVs, but its signal to the other UAV is blocked by a building. 

Information communicated between the ground user and the UAVs is either backhauled to some central gateway when the destination is external to the network, or it is routed in the air to another UAV when the destination is internal to it.    Backhauling and routing are outside the scope of this paper, but rather we focus exclusively on the UAV-to-ground link.

The city is laid out as a grid with streets either going in the east-west direction (X-axis) or the north-south direction (Y-axis).  The locations of the streets are drawn from a MPLP, which is a pair of independent one-dimensional Poisson point processes (PPPs) of density $\lambda_\mathsf{s}$, one for each dimension along the ground plane.  In each dimension, the interval between neighboring points corresponds to the width of one street and the length of one building block.  The mean street width is $\mu_\mathsf{s}$ and the mean block length is $\mu_\mathsf{b}$. Accordingly, $\lambda_\mathsf{s}=1/(\mu_\mathsf{s}+\mu_\mathsf{b}).$    It is assumed that each block contains at least one building, and if there are multiple buildings, they are all the same height.   The height of a block is represented by the random variable $H_B$ with cumulative distribution function $F_{H_B}(h)$ and mean $\mu_H$.  For ease of exposition, we assume that $H_B$ is uniformly distributed between a minimum height $H_\mathsf{min}$ and a maximum height $H_\mathsf{max}$, but other distributions such as exponential or Rayleigh can be considered (see \cite{Heath20211}).

Let $Y$ denote a reference user and its location. Without loss of generality, assume that the user is located at the origin; i.e., $Y=0$.  The user is in a vehicle with an antenna of height $h_\mathsf{V}$ and located either on a north-south street of width $w_v$, on an east-west street of width $w_h$, or at an intersection, in which case it is simultaneously on a north-south street and an east-west street.   Accordingly, we consider two distinct cases: (1) The vehicle is at an intersection, or (2) The vehicle is not at an intersection.   Note that under this model, the reference station is at a fixed location while the topology of the city (location of streets and heights of buildings) changes for each realization of the underlying random model.  Clearly, this is not literally how a city behaves -- the buildings and streets are in known locations and do not change frequently.  However, for a large city with a homogeneous distribution of streets and buildings, this is equivalent to randomly dropping the station at a different location within the city during each realization of the model. Thus, this is an effective model for a moving or randomly placed user and enables a good characterization of the city as a whole.

\begin{figure}[t]
		\centering
		\includegraphics[width=.81\linewidth] {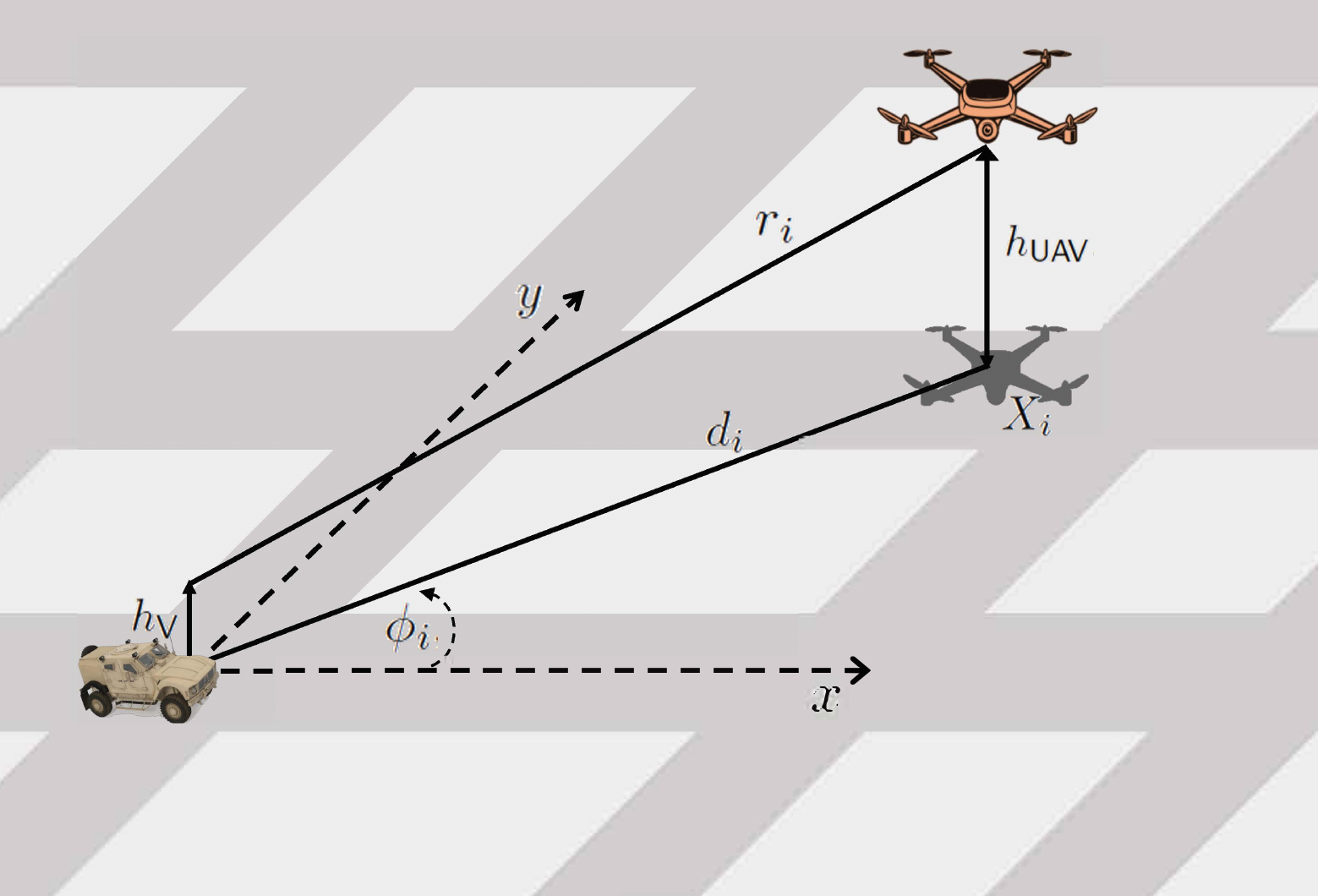}
	    \caption{  Coordinates associated with the $i^\mathsf{th}$ UAV, including its location $X_i$ as projected onto the ground, the 2D distance $d_i$ from the vehicle to the projection, the azimuth angle $\phi_i$ relative to the positive $X$ axis,  the height of UAV $h_{\mathsf{UAV}}$, the height of the vehicle antenna $h_\mathsf{V}$, and the 3D distance $r_i$. \vspace{-.5cm} }
		\label{fig:Model3D4}
	\end{figure}

We assume that the UAVs are randomly deployed and model their locations with a two-dimensional homogeneous PPP with density $\lambda_{\mathsf{UAV}}.$  The UAVs are all at a constant height (or altitude) $h_{\mathsf{UAV}}.$   As illustrated in Fig.~\ref{fig:Model3D4}, let $X_i$ for i $\in {\mathbb Z}^+$ denote the UAVs and their locations as projected onto the ground plane. Let $d_i = |X_i|$ be the distance from $Y$ to $X_i$. As shown in Fig.~\ref{fig:Model3D4},  $d_i$  is the distance from the user's vehicle to the projection of the UAV onto the ground (henceforth called the ``2D distance''), while $r_i$ is the distance from the vehicle's antenna to the UAV (the ``3D distance''). The azimuth angle of departure of the UAV-vehicle path is denoted by $\phi_{i}$,  which is uniform over $(0,2\pi)$ since the UAVs are drawn from a homogenous PPP. The UAVs are indexed according to their distances to $Y$ such that  $d_1 \leq d_2\leq....$ 

The user may connect to any unblocked UAV that is within a maximum 3D  transmission range $r_{\mathsf{max}}$.   The maximum range $r_{\mathsf{max}}$ accounts for physical-layer parameters including the transmit power, antenna gains, noise floor, fading margin, atmospheric effects, and transmission technology.  To account for the UAVs being above the ground plane, the maximum 3D transmission range must be projected onto a maximum 2D transmission range $d_\mathsf{max}=\sqrt{r_{\mathsf{max}}^2-(h_{\mathsf{UAV}}-h_{\mathsf{V}})^2}$.   The user is considered to be \emph{connected} if it has a LOS path to any UAV whose ground projection $X_i$ is within the disk of radius $d_\mathsf{max}$. Because the UAVs are drawn from a PPP, it follows that the number $N$ of UAVs within this disk is Poisson with mean $ \lambda_{\mathsf{UAV}} \pi d_{\mathsf{max}}^2$.  The set of all UAVs within this disk forms a set of prospective UAVs with which the user may communicate.  Denote this set by $\mathcal X$; i.e., $\mathcal{X}=\{X_1,X_2,...,X_N\}.$

\section{LoS Distribution and Outage Probability}


The \emph{connectivity probability} $p_{\mathsf{c}}$ is the probability that at least one UAV within the maximum transmission range has an unblocked LoS to the user.  The connectivity probability depends on whether the user's vehicle is at an intersection or outside an intersection.  Initially, assume that it is at an intersection (the alternative case will be considered later). The connectivity probability also depends on the UAV locations $\mathcal{X}$ as well as the realization of the city environment (i.e., the MPLP governing the street locations and the building heights). First, assume the UAVs locations $\mathcal{X}$ are fixed (this conditioning will be removed later). The corresponding \emph{conditional} connectivity probability is denoted by $p_{\mathsf{c}}(\mathcal{X})$. 


In an urban environment, there is likely to be at least some correlation in the blockage events, since a large building could potentially block many UAVs.   However, for an initial analysis, we assume here that the UAVs are independently blocked.  Extension of this analysis to correlated blocking is left for future study as we discuss in the Conclusion.
Under the assumption of independent blocking and noting that the vehicle is unable to connect only when \emph{all} UAVs within the maximum range are blocked (i.e., non-LoS), the conditional probability of coverage can be found as
\begin{eqnarray}
p_{\mathsf{c}}(\mathcal{X})=1-\prod_{i=1}^{N} \left( 1-p_{\mathsf{LoS}}\left(X_i\right) \right)  \label{pcoverall}
\end{eqnarray} 
where $p_{{\mathsf{LoS}}}(X_i)$ is the probability that the path between $Y$ and $X_i$ is LoS. 

As in \cite{Heath20211}, the value of $p_{\mathsf{LoS}}(X_i)$ may be found by noting that (1) The intersection containing the vehicle is at a known location (the origin), and hence the pair of perpendicular streets flowing through that intersection are at deterministic locations, and (2) The other streets are drawn from the MPLP, which comprises two \emph{independent} PPPs, and hence the north-south and east-west streets are independent.   These properties allow $p_{\mathsf{LoS}}(X_i)$ to be decomposed as
\begin{eqnarray}
p_{\mathsf{LoS}}(X_i)= p_{\mathsf{LoS}}^{(0)}(X_i)~ p_{\mathsf{LoS}}^{(x)}(X_i)~ p_{\mathsf{LoS}}^{(y)}(X_i)
\label{eq:plos2}
\end{eqnarray} 
where $p_{\mathsf{LoS}}^{(0)}(X_i)$ is the probability that the link is not blocked by the building closest to the vehicle, which is at a deterministic location and on the same street as the vehicle, $p_{\mathsf{LoS}}^{(x)}(X_i)$ is the probability that it is not blocked by a building side that is perpendicular to the X-axis (governed by the first PPP), and $p_{\mathsf{LoS}}^{(y)}(X_i)$ is the probability that it is not blocked by a side that is perpendicular to the Y-axis (governed by the second PPP).

Finding the first factor in (\ref{eq:plos2}) requires identifying a critical height $h^{(0)}_i$ for the building.  If the building neighboring the intersection is taller than this height, then $X_i$ will be blocked by it.  From the geometry of Fig. \ref{fig:Model3D4} (see also  \cite{Heath20211}), this height is
\begin{eqnarray}
h^{(0)}_i&=&\frac{\mathsf{max}\left( \frac{w_\mathsf{v}}{2} ,\frac{w_\mathsf{v}}{2}~ \mathsf{cot}\phi_i \right)\left(h_{\mathsf{UAV}}-h_\mathsf{V}\right)}{d_i \mathsf{cos}\phi_i}+h_\mathsf{V}.
\end{eqnarray}
It follows that the probability that the nearest building does not block the link is the CDF of the building height evaluated at the critical height
\begin{eqnarray}
p_{\mathsf{LoS}}^{(0)}(X_i)=F_{{H}_B}\left(h^{(0)}_i\right).
\end{eqnarray}
%

The second and third factors of (\ref{eq:plos2}) require determining the likelihood that there is a building side in the corresponding direction that is sufficiently high to block the UAV.   If building heights were ignored, then the probability can be found from the void probability of the corresponding PPP, which gives the probability that there are no points (building sides) in the desired interval.  For a homogeneous PPP of intensity $\lambda$ and interval of length $\ell$, the void probability is $\exp (-\lambda \ell)$.   However, when height is considered, the PPP is thinned such that any point whose corresponding building side has a height that is less than the critical height for that location must be removed.  The result is a non-homogenous PPP whose density is a function of distance, and its void probability is found by integrating the density along the interval.  Letting $Z \in \{ X, Y \}$ indicate either the X or Y direction, the resulting void probability of the thinned PPP is
\begin{eqnarray}
   p_{\mathsf{LoS}}^{(z)}(X_i)
   & = & 
   \exp \left( 
   - \int_{z_\mathsf{a}}^{z_\mathsf{b}} \lambda^{(z)}_\mathsf{s}(z) dz
   \right)
   \label{nonhomogeneous}
\end{eqnarray}
where $\lambda^{(z)}_\mathsf{s}(z)$ is the non-homogenous density of the thinned PPP in the Z direction, while $z_\mathsf{a}$ and $z_\mathsf{b}$ are the limits of the integration (discussed below).

A point at coordinate $z$ will be removed if the corresponding building side at that coordinate is below a critical height for that coordinate, otherwise it will be retained.  Since the probability that a point at coordinate $z$ will be retained is the complement of the CDF of the building height evaluated at that coordinate, the density of the thinned PPP becomes
\begin{eqnarray}
   \lambda^{(z)}_\mathsf{s}(z)
   & = & 
   \lambda_\mathsf{s} 
   \left[1-F_{{H}_B} \left( h^{(z)}_i(z) \right)\right]
\end{eqnarray}
where $h^{(z)}_i(z)$ is the critical height in the Z direction at coordinate z, which can be found from the geometry of the problem.

In the X direction, the critical height is 
\begin{eqnarray}
h^{(x)}_i(x)
& = &
\frac{x\left(h_{\mathsf{UAV}}-h_\mathsf{V}\right)+h_\mathsf{V} d_i \mathsf{cos}\phi_i}{d_i \mathsf{cos}\phi_i}
\end{eqnarray}
while in the Y direction it is
\begin{eqnarray}
h^{(y)}_i(y) 
& = & 
\frac{y\left(h_{\mathsf{UAV}}-h_\mathsf{V}\right)+h_\mathsf{V} d_i \mathsf{sin}\phi_i}{d_i \mathsf{sin}\phi_i}. 
\end{eqnarray} 

For the X direction, the limits of integration in (\ref{nonhomogeneous}) are
\begin{eqnarray}
z_\mathsf{a} & = & \mathsf{max}\left( \frac{w_\mathsf{v}}{2}, \frac{w_\mathsf{v}}{2} \mathsf{cot}\phi_i \right) \nonumber \\
z_\mathsf{b} & = & d_i \mathsf{cos}\phi_i
\end{eqnarray}
while for the Y direction they are
\begin{eqnarray}
z_\mathsf{a} & = &\mathsf{max}\left( \frac{w_\mathsf{h}}{2}, \frac{w_\mathsf{h}}{2}\mathsf{tan}\phi_i \right) \nonumber \\
z_\mathsf{b} & = & d_i \mathsf{sin}\phi_i.
\end{eqnarray}

Since the connectivity probability depends on whether or not the vehicle is at an intersection, we define $p_{{\mathsf{c}}_{\mathsf{sec}}}(\mathcal{X})$ to be the conditional connectivity probability when the vehicle is at an intersection and $p_{{\mathsf{c}}_{\mathsf{str}}}(\mathcal{X})$ to be the conditional connectivity probability when it is not. $p_{{\mathsf{c}}_{\mathsf{sec}}}(\mathcal{X})$ is found using the methodology given above, while $p_{{\mathsf{c}}_{\mathsf{str}}}(\mathcal{X})$ is found by setting $w_\mathsf{v}=0$ or $w_\mathsf{h}=0$ (depending on whether the vehicle is located on a north-south street or an east-west street).




Because the UAVs are randomly located, the value of $p_{\mathsf{c}}(\mathcal{X})$ will vary from one network realization to the next. Hence $p_{\mathsf{c}}$ is itself a random variable, and the CDF of $p_{\mathsf{c}}$ can be described as
\begin{eqnarray}
F_{\mathsf{c}}(\gamma)= \mathbb{P}[p_{\mathsf{c}}\leq \gamma].
\label{eq:CDFLoS1}
\end{eqnarray} 

The overall CDF is the average of the CDFs of $p_{{\mathsf{c}}_{\mathsf{sec}}}$ and $p_{{\mathsf{c}}_{\mathsf{str}}},$ where the probability of being in an intersection is $\mu_\mathsf{s}/(\mu_\mathsf{s}+\mu_\mathsf{b})$. Equation (\ref{eq:CDFLoS1}) can be rewritten as
\begin{eqnarray}
F_{\mathsf{c}}(\gamma)&=&\left(\frac{\mu_\mathsf{s}}{\mu_\mathsf{s}+\mu_\mathsf{b}}\right) \mathbb{P}[p_{{\mathsf{c}}_{\mathsf{sec}}}\leq \gamma]  \nonumber \\
 & & +\left(1-\frac{\mu_\mathsf{s}}{\mu_\mathsf{s}+\mu_\mathsf{b}}\right)\mathbb{P}[p_{{\mathsf{c}}_{\mathsf{str}}}\leq \gamma]. 
\label{eq:CDFLoS2}
\end{eqnarray}

The CDF of $p_\mathsf{c}$, $F_{\mathsf{c}}(\gamma)$, quantifies the likelihood that the  $p_{\mathsf{c}}$ is below some threshold $\gamma$.  If~$\gamma_\mathsf{th}$ is interpreted as the minimum acceptable $p_{\mathsf{c}}$ required to achieve reliable communications, then $F_{\mathsf{c}}(\gamma_\mathsf{th})$ is the \emph{outage probability} of the system. 
The outage probability can be used to evaluate the tradeoff of using different UAV densities. In addition, it can be used to find the optimal UAV height for a given UAV density and city type.  The optimization of UAV height may be expressed as
\begin{equation}
\begin{aligned}
\min_{h_{\mathsf{UAV}}} ~  F_{\mathsf{c}}(\gamma_\mathsf{th}), \quad
\textrm{s.t.} \quad & 0 < r_{\mathsf{max}}+h_{\mathsf{V}} < h_{\mathsf{UAV}}\\
\end{aligned}
\label{eq:Opti}
\end{equation}
While (\ref{eq:Opti}) could be solved using classical gradient descent approaches, a pragmatic approach to solving it is to simply generate the $p_{\mathsf{c}}$ distribution for a range of UAV densities and heights, read the outage probability from those curves, and look for the height that minimizes the outage probabilty for each UAV density.

\vspace{-0.1cm}

\section{Numerical Results}

\begin{table}
\centering
\caption{\sc Parameters for the Urban Grid Model}
\scriptsize
\begin{tabular}{ |c|c|c|c|c| } 
\hline
\multirow{2}{4em} {Type} & Mean building
 & Mean side
 & Mean street
 \\
 &height, $\mu_\mathsf{H}$ & width, $\mu_\mathsf{b}$& width, $\mu_\mathsf{s}$\\
 \hline\hline
Suburban & 10 m & 37 m & 10 m \\ 
Urban & 19 m & 45 m & 13 m \\
Dense urban & 25 m & 60 m & 20  m \\
\hline
\end{tabular}
\vspace{-0.2cm}	
\end{table} 
 
 \begin{figure}[t]
	\centering
	\includegraphics[width=0.9\textwidth]{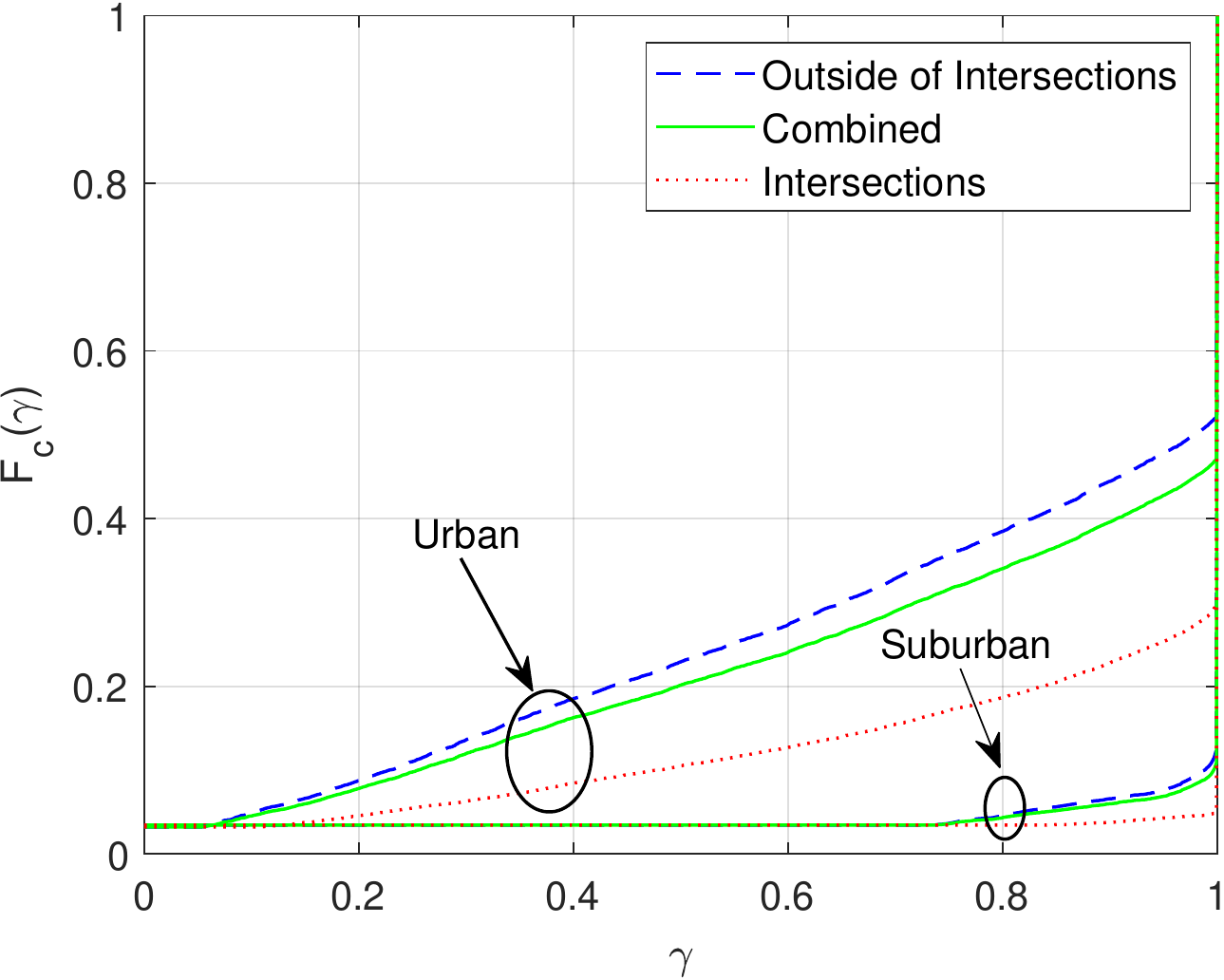}
	\caption{The connectivity distribution for urban and suburban deployments, considering two possible vehicle locations (at intersections or outside intersections) and the average across both locations. UAV density, UAV height, and maximum transmission range are fixed at $\lambda_\mathsf{UAV}=20/\mathsf{km^2}$, $h_\mathsf{UAV}=100$ m, and $r_\mathsf{max}=250$ m.\vspace{-0.6cm}}
	\label{fig:CDF_pLoS_Different_Grid_DiffScenario}

\end{figure}

In this section, we begin by illustrating the connectivity distribution through numerical examples.   We then investigate the outage probability for a variety of city types, UAV densities, and UAV heights.  Finally, we find the optimal UAV altitude that minimizes the outage probability for different city types as a function of UAV density.    Three city types are considered: Suburban, urban, and dense urban.  The parameters used for each of these city types is specified in Table I.  When non-zero values are needed for $w_\mathsf{h}$ and $w_\mathsf{v}$, they are set to the value of $\mu_\mathsf{s}$ given in the table.   The minimum and maximum building heights are $H_\mathsf{min} = \mu_H/2$ and $H_\mathsf{max} = 3\mu_H/2$, respectively. Throughout this section, the height of the vehicle's antenna is set to $h_\mathsf{V} = 10~\mathsf{m}$.

Because  $p_\mathsf{c}$ depends on $\mathcal{X}$, its distribution $F_\mathsf{c}(\gamma)$ can be found by defining the equivalent event $\mathcal X' = \left\{ \mathcal X : p_\mathsf{c}( \mathcal X) \leq \gamma \right\}$ and then finding the probability that $\mathcal X$ is in $\mathcal X'$; i.e., $F_\mathsf{c}(\gamma) = \mathbb P ( \mathcal X \in \mathcal X')$.   However, this requires the inversion of the equation for $p_\mathsf{c}(\mathcal X)$, which is not a tractable operation, especially since the cardinality of $\mathcal X$ (i.e., $N$) is random.  Moreover, the integral required to obtain the probability is over a region with an irregular boundary.   Instead,  an effective and pragmatic approach to finding the distribution of $p_\mathsf{c}$ is to use a simulation.   The simulation involves the repeated realization of the set $\mathcal{X}$. For each realization of $\mathcal{X}$, the resulting $p_\mathsf{c}(\mathcal X)$ is found and recorded.  The CDF of $p_\mathsf{c}$  is then found by plotting the empirical CDF of the recorded data.

The connectivity distribution $F_\mathsf{c}(\gamma) = \mathbb P [ p_\mathsf{c} \leq \gamma]$ is shown in Fig.~\ref{fig:CDF_pLoS_Different_Grid_DiffScenario} for urban and suburban environments, taking into account the two kinds of vehicle locations. In particular, the blue dashed lines correspond to the case that the vehicle is located at an intersection, the dotted red lines correspond to the case that the vehicle is located outside an intersection, and the green solid lines account for both cases by averaging them according to (\ref{eq:CDFLoS2}). The distributions are computed by fixing the values of $r_\mathsf{{max}}=250~\mathsf{m},$ $\lambda_\mathsf{UAV}=20/\mathsf{km^2},$ and $h_\mathsf{UAV}=100~\mathsf{m}.$ 

When comparing the connectivity distribution with figures such as Fig.~\ref{fig:CDF_pLoS_Different_Grid_DiffScenario}, curves that are relatively lower correspond to better performing scenarios than curves that are higher.  This is because the lower curves are less likely to have coverage probabilities that fail to meet a threshold (the value of the argument $\gamma$), and hence are \emph{more} likely to have coverage probabilities that are better than $\gamma$.  Thus, Fig.~\ref{fig:CDF_pLoS_Different_Grid_DiffScenario} shows that performance is consistently better in suburban environments than in urban environments, which can be attributed to the lower value of $\mu_\mathsf{H}$ and hence shorter buildings in the suburbs.  The figure also shows that performance is better when the vehicle is at intersections than when it is not at an intersection.  This is due to the fact that, at an intersection, the vehicle can find a LoS path directly down two streets instead of just down the one street.  When looking at the combined probability, we see that the performance is closer to the case where it is outside an intersection, since a vehicle moving at constant velocity is more likely to be outside of an intersection than it is to be at an intersection.  For the remaining results in this section, the connection probability is found by averaging the probabilities at the two vehicle locations.


 \begin{figure}[t]
	\centering
	\includegraphics[width=0.92\textwidth]{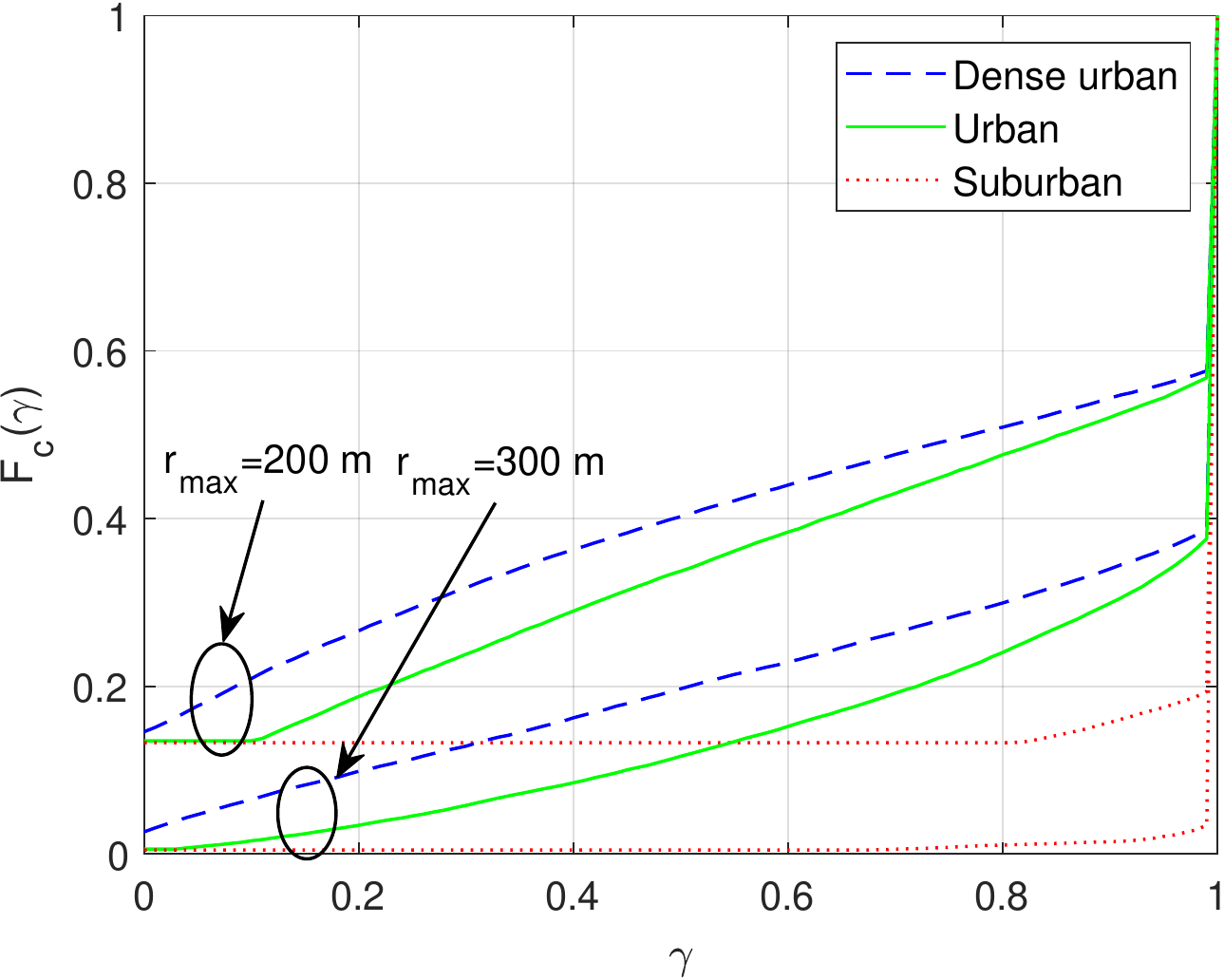}
	\caption{The connectivity distribution when $r_\mathsf{max}={200}$ m and $300$ m for dense urban, urban, and suburban deployments. UAV density and UAV height are fixed at values of $\lambda_\mathsf{UAV}=20/ \mathsf{km^2}$, $h_\mathsf{UAV}=100~\mathsf{m}.$}
	\label{fig:CDF_pLoS_Different_R_DiffGrdiGeometry}
\end{figure}

Fig.~\ref{fig:CDF_pLoS_Different_R_DiffGrdiGeometry} shows the connectivity distribution for two different values of the maximum transmission range, $r_\mathsf{max} = \{200,300\}~\mathsf{m}$, taking into account all three different city types.  When $r_\mathsf{max}$ is greater, the CDF is lower, due to the fact that the vehicle is likely to have more UAVs within its range. The CDF also varies depending on the deployment, since building heights are generally higher for urban deployments and even higher for dense urban deployments, hence the denser environments perform worse than the sparser ones. 

 

\begin{figure}[t]
	\centering
	\includegraphics[width=0.92\textwidth]{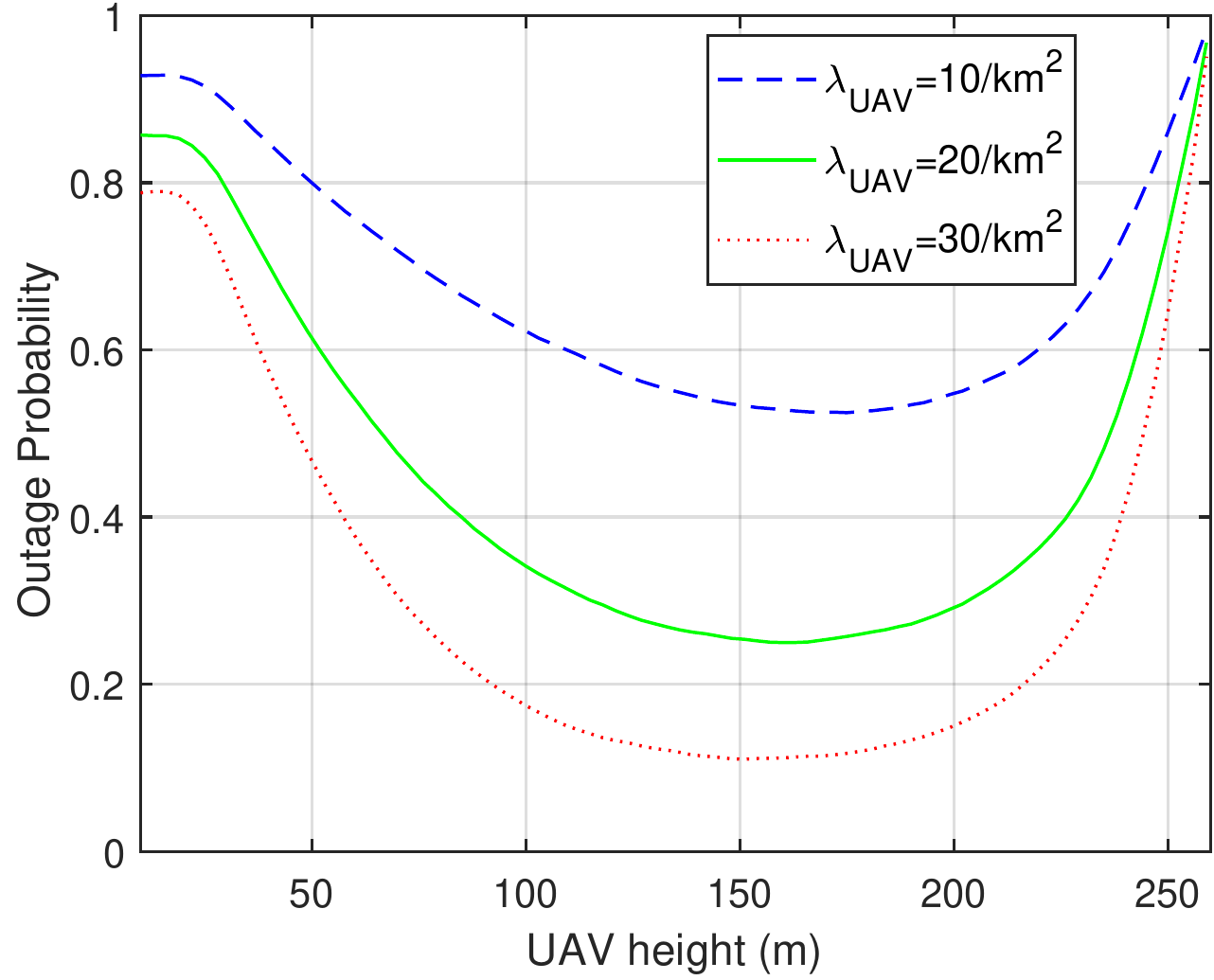}
	\caption{The outage probability at threshold $\gamma_\mathsf{th}=0.8$ with respect to $h_\mathsf{UAV}$ when $\lambda_\mathsf{UAV}={10,20,30}/ \mathsf{km^2}$. The deployment is urban and the maximum transmission range is fixed at  $r_\mathsf{max}=250~\mathsf{m}.$ \vspace{-0.4cm}}
	\label{fig:Outage_vshUAV_Difflambda}	
\end{figure}

\begin{figure}[t]
\centering
	\includegraphics[width=0.92\textwidth]{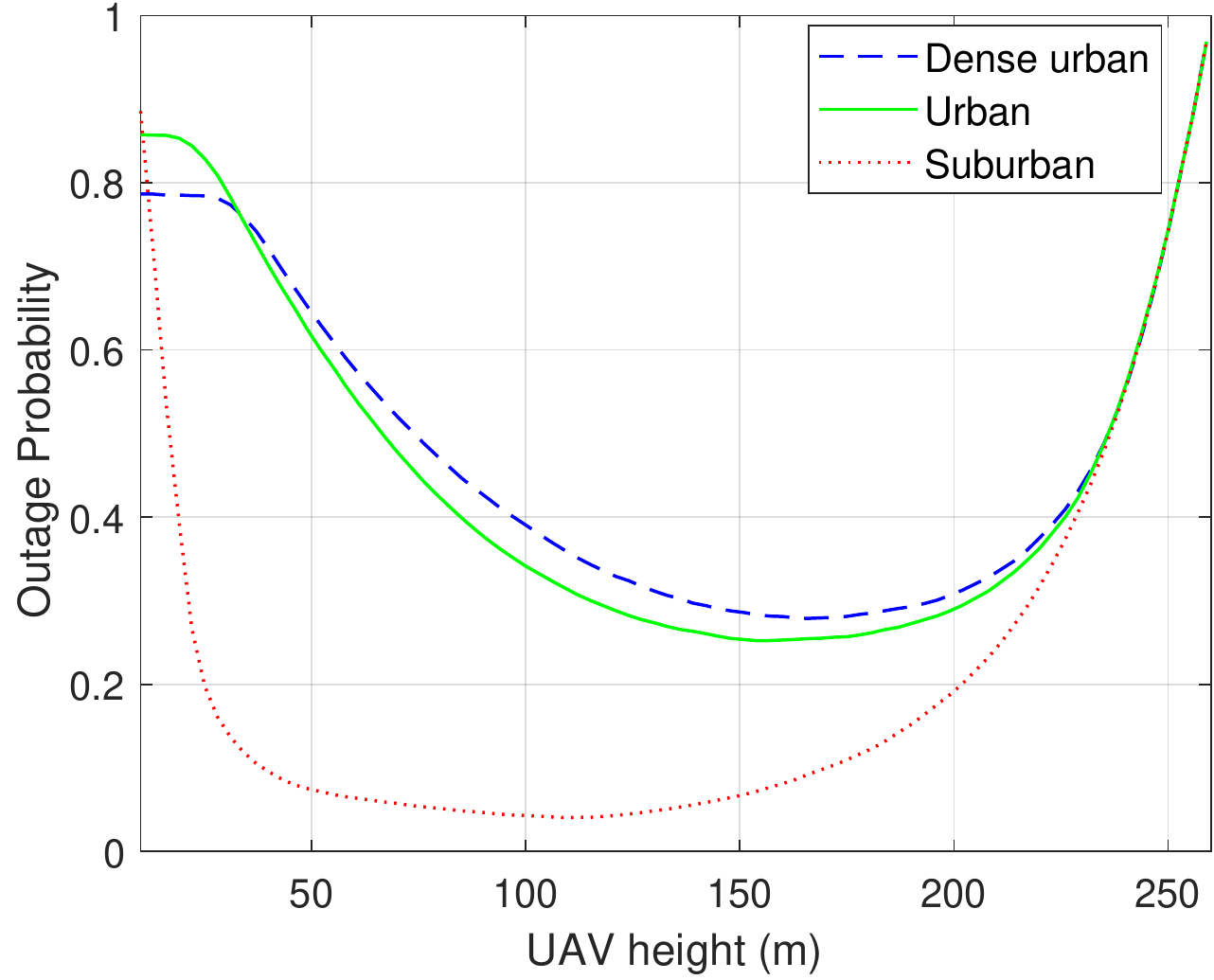}
	\caption{The outage probability at threshold $\gamma_\mathsf{th}=0.8$ with respect to $h_\mathsf{UAV}$ for the three city types. UAV density and maximum transmission range are fixed at values of $\lambda_\mathsf{UAV}=20/ \mathsf{km^2}$ and $r_\mathsf{max}=250~\mathsf{m}.$}
	\label{fig:Outage_vshUAV_DiffGridGeometry}
\end{figure}



	\begin{figure}[t]
\centering
	\includegraphics[width=0.92\textwidth]{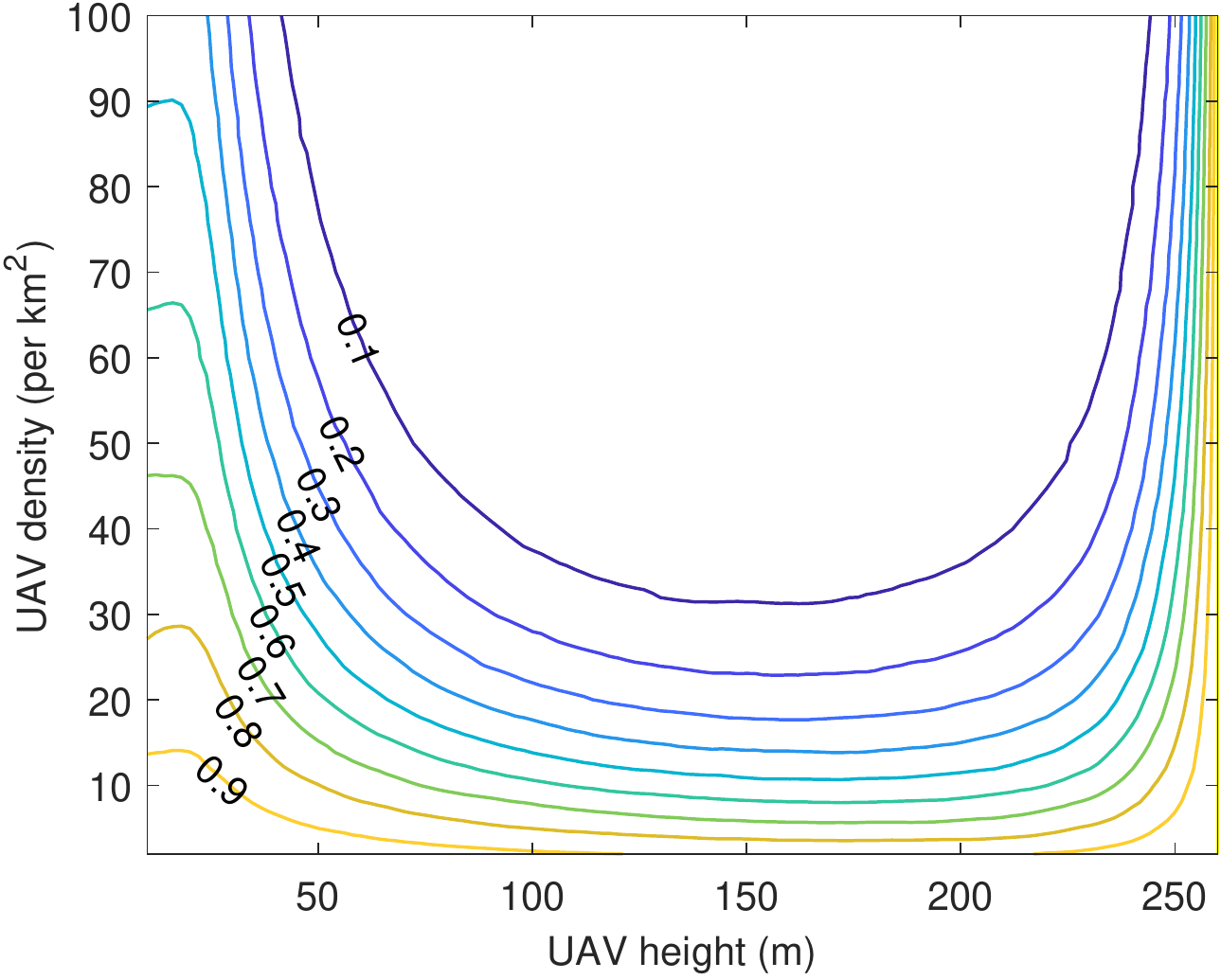}
	\caption{Contour plot showing the outage probability as a function of both UAV density $\lambda_\mathsf{UAV}$ and UAV height $h_\mathsf{UAV}$.  Contours are spaced in increments of 0.1, with 0.1 being the topmost curve. \vspace{-0.4cm}
}
	\label{fig:Contour}
	
\end{figure} 


Figs.~\ref{fig:Outage_vshUAV_Difflambda} and \ref{fig:Outage_vshUAV_DiffGridGeometry} show the outage probability rather than the connectivity distribution.  The outage probability is merely the value of the connectivity distribution at a specified threshold $\gamma_\mathsf{th}$.   Hence, the outage probability is $F_{\mathsf{c}}(\gamma_\mathsf{th})$ and quantifies the probability that a given network realization fails to have a coverage probability of $\gamma_\mathsf{th}$. For these curves, 
 $\gamma_\mathsf{th} = 0.8$.   Fig.~\ref{fig:Outage_vshUAV_Difflambda} shows the outage probability as a function of UAV height in an urban environment for three different values of UAV density $\lambda_\mathsf{UAV} = \{10, 20, 30\}/ \mathsf{km^2}$ with $r_\mathsf{max} = 250~\mathsf{m}$.   The figure shows that there is an optimal value of the UAV height $h_\mathsf{UAV}$.  If  $h_\mathsf{UAV}$ is too low, it will be more easily blocked by buildings that are tall enough to intersect the LoS path.  However, if $h_\mathsf{UAV}$ is too high, the projection $d_\mathsf{max}$ onto the ground plane will get smaller due to the geometry (see Fig. 2), and thus the average number of UAVs within range will be reduced.  The optimal height increases with decreasing UAV density, suggesting that when the UAV density is low, it is more important for close UAV to clear nearby buildings by being sufficiently high than it is for there to be a large area containing UAVs.

 
Fig.~\ref{fig:Outage_vshUAV_DiffGridGeometry} shows the outage probability as a function of UAV height for all three city types at UAV density $\lambda_\mathsf{UAV} = 20 / \mathsf{km^2}$ with $r_\mathsf{max} = 250~\mathsf{m}$.   The figure shows that the optimal height depends on the type of city environment, since denser city environments have a taller mean building height.   As the city environment gets denser, higher UAV altitudes are required to overcome blocking by the taller buildings.  

Fig.~\ref{fig:Contour} is a contour plot that shows contours of constant outage probability as a function of both UAV density $\lambda_\mathsf{UAV}$ and UAV height $h_\mathsf{UAV}$.   Contours are spaced in increments of 0.1, with 0.1 being the top curve and 0.9 the bottom curve.   The outage threshold is $\gamma_\mathsf{th} = 0.8$, the city type is urban, and the maximum transmission range is $r_\mathsf{max} = 250~\mathsf{m}$.  The figure can be used to determine the minimum UAV density and optimal height required to meet an outage constraint.  For instance, if the system must operate at an outage of 0.1, then the curve shows that the UAV density must be at least $\lambda_\mathsf{UAV} = 31 / \mathsf{km^2}$ and the height should be $h_\mathsf{UAV} = 162~\mathsf{m}$ when the minimum UAV density used.



\begin{figure}[t]
	\centering
	\includegraphics[width=0.9\textwidth]{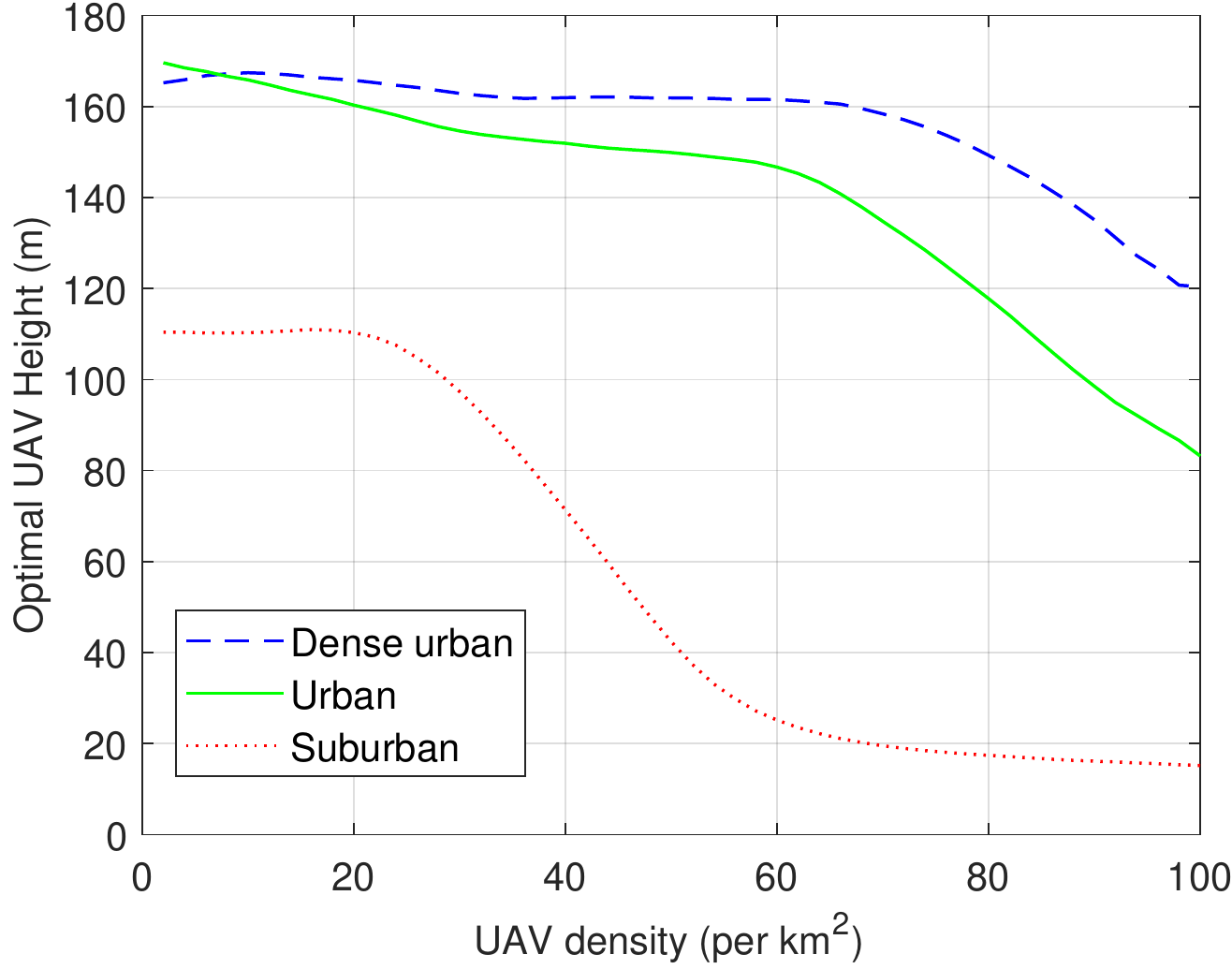}
	\caption{The optimal UAV height with respect to $\lambda_\mathsf{UAV}$ for dense urban, urban, and suburban deployments. \vspace{-0.4cm}}
	\label{OptimalHeight_lambda}
\end{figure}

\begin{figure}[t]
	\centering
	\includegraphics[width=0.9\textwidth]{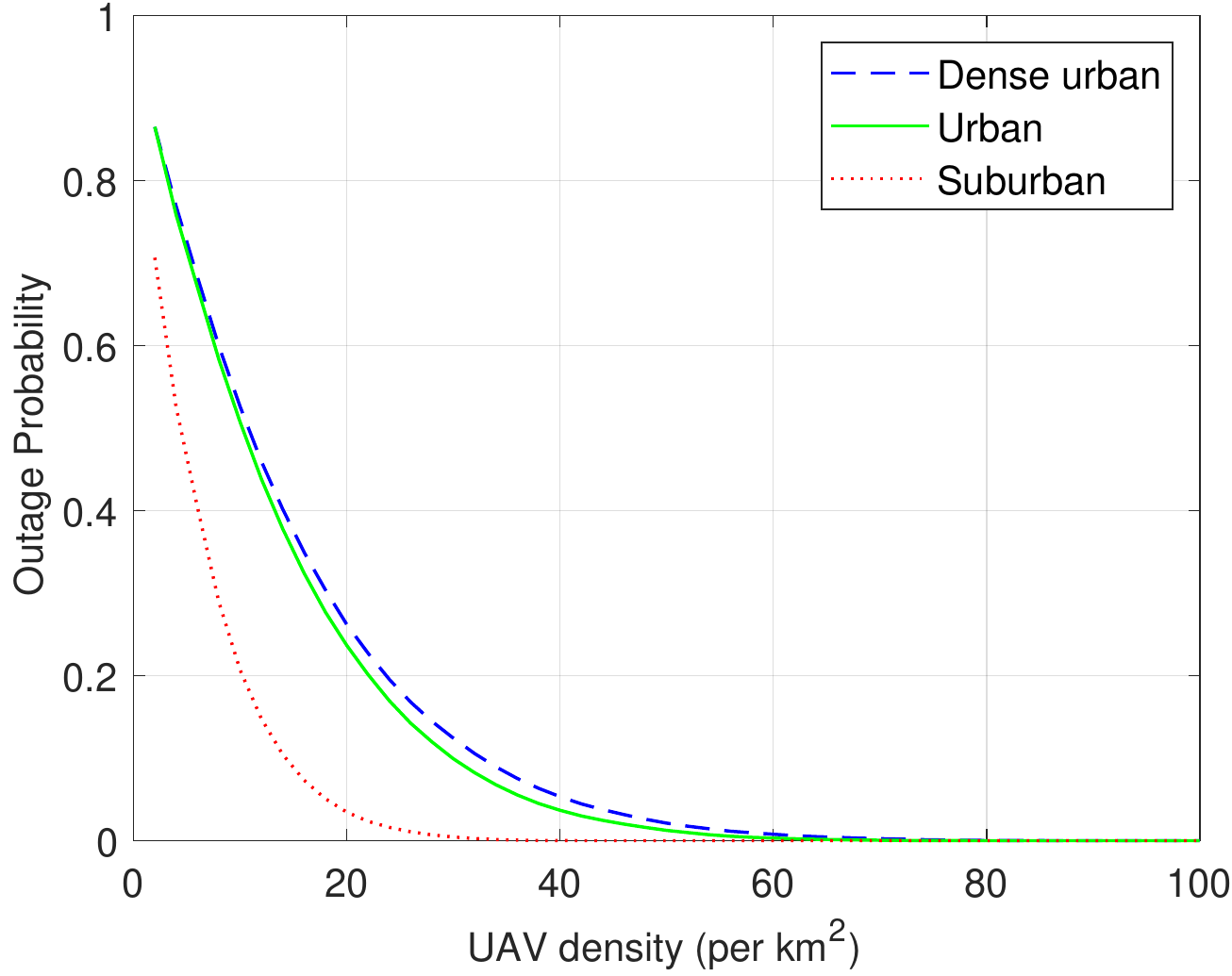}
	\caption{Outage probability achieved when the optimal UAV height is used in dense urban, urban, and suburban deployments.\vspace{-0.4cm}}
	\label{fig:OutageAtOptimalHeight_vs_lambda1}
\end{figure}

Figs.~\ref{OptimalHeight_lambda} and \ref{fig:OutageAtOptimalHeight_vs_lambda1} show the result of the optimization given by (\ref{eq:Opti}) for a threshold of $\gamma_\mathsf{th} = 0.8$ and $r_\mathsf{max} = 250~\mathsf{m}$. In particular, Fig.~\ref{OptimalHeight_lambda} shows the optimal UAV height as a function of UAV density for each of the three city types, and Fig.~\ref{fig:OutageAtOptimalHeight_vs_lambda1} shows the corresponding outage probability.  From these curves, it is again observed that the optimal height decreases with increasing UAV density and that the optimal height is lower for less dense city environments.  Together, these curves can be used to determine the minimum UAV density required to achieve a desired outage probability (by reading Fig.~\ref{fig:OutageAtOptimalHeight_vs_lambda1}) and the corresponding optimal UAV height (by reading from Fig.~\ref{OptimalHeight_lambda}).

\section{Conclusion}

By using a MPLP to model the locations of city streets and taking into account the random heights of the buildings, the methodology in this paper can be used to compute the connectivity probability for a randomly deployed network of multiple UAVs.  For UAVs at fixed location, the conditional connectivity probability characterizes the connectivity conditioned on those locations.  When the UAVs are drawn from a point process, then removing the conditioning allows the connectivity to be characterized by its distribution, which gives the probability that a network realization (set of UAV locations) fails to meet a threshold.   The connectivity distribution can then be used to find the outage probability, which is the distribution evaluated at a specified outage threshold.    With the outage probability so defined, it can be used to characterize the tradeoff between UAV density and network performance for different city types.   Moreover, the outage probability can be used to drive an optimization that identifies the optimal UAV height.



We note some limitations of this work that imply directions for future work.  We assume that blocking is independent, but in reality it is likely that a large building may block several UAVs simultaneously.  The extent of this issue could be initially explored through simulation, and if correlated blocking proves to significantly impact performance, then the analysis should be extended to account for it.   Such an extension is beyond the scope of this paper, but could potentially avail of the approximations provided in \cite{Heath2019}.  With respect to optimization, we assume that all UAVs are at a constant altitude, but it may prove to be beneficial for UAVs to be at different altitudes.  Instead of optimizing a fixed altitude, the optimization could be used to identify an optimizing altitude distribution.   Finally, we note that the physical-layer model is captured by a single parameter -- the maximum transmission range -- but a more detailed analysis could explicitly capture the effects of fading, interference, and antenna beamforming. Such analysis could also capture the effects of building reflections.

While this paper has focused on mmWave transmission, it applies equally well to other modes of communication that are primarily LoS, such as free-space optical (FSO) communication.

\section*{Acknowledgements}
Heath was supported in part by the National Science Foundation under Grant No. ECCS-1711702 and Valenti was supported in part by the National Science Foundation under Grant No. CNS-1066197.
\balance

\small
\bibliographystyle{ieeetr}

\balance
\bibliography{./References}

\end{document}